\documentclass[showpacs,preprintnumbers,amsmath,amssymb,superscriptaddress]{revtex4}

\usepackage{color,epsfig} 
\usepackage{graphicx}

\def\be{\begin{equation}}
\def\ee{\end{equation}}

\renewcommand{\baselinestretch}{1.5}

\begin{document}
\preprint{}

\title{\bf SCALING DYNAMICS OF DOMAIN WALLS \\ IN THE CUBIC ANISOTROPY MODEL}
\author{Richard A. Battye}
\affiliation{Jodrell Bank Observatory, University of Manchester, Macclesfield, Cheshire SK11
9DL, UK.}
\author{Adam Moss}
\affiliation{Jodrell Bank Observatory, University of Manchester, Macclesfield, Cheshire SK11
9DL, UK.}

\date{\today}
 
\begin{abstract}
We have investigated the dynamics of domain walls in the cubic anisotropy model. In this model a global $O(N)$ symmetry is broken to a set of discrete vacua either on the faces, or vertices of a (hyper)cube. We compute the scaling exponents for $2\le N\le 7$ in two dimensions on grids of $2048^2$ points and compare them to the fiducial model of $Z_2$ symmetry breaking. Since the model allows for wall junctions lattice structures are locally stable and modifications to the standard scaling law are possible. However, we find that since there is no scale which sets the distance between walls, the walls appear to evolve toward a self-similar regime with $L\sim t$.
\end{abstract}


\maketitle

\section{Introduction}

The breaking of spontaneous symmetries in the early universe can produce topological defects which can affect the cosmological evolution. Over the years defect networks have received much attention - for a detailed review see~\cite{HK,VS}. Most of this work has focused on cosmic strings - the study of domain walls has generally been discouraged due to the commonly accepted view that such a network will either over-close the universe or destroy the isotropy of the Cosmic Microwave Background (CMB)~\cite{ZKO}. However, this assumption relies on the fact that the network immediately enters a self-similar scaling regime and the walls are formed at a high energy transition. In this scenario, wall decay processes are efficient and, on average, there is a single domain wall per horizon volume from initial formation through to the present day with domains growing as fast as causality allows (the characteristic lengthscale $L\sim t$). This behaviour has been confirmed numerically for models where the potential of a scalar field has $Z_{2}$ symmetry and there are two vacua~\cite{RPS,CLO,LSW,GH,AM1,AM2}. An alternative kind of behaviour has been envisaged for models with more complicated vacuum structure. It could be that the network frustrates and $L\sim a(t)$\cite{V,BS,BBS} allowing the network to act as a dark energy candidate with pressure to density ratio $w=P/\rho=-2/3$, if the scale of symmetry breaking is sufficiently low.

In this paper, we investigate the stability properties and dynamical evolution of a network of domain walls where a global $O(N)$ symmetry is broken to a set of discrete vacua with (hyper)cubic symmetry. In these models walls can be joined at string-like junctions in 3D (vortex-like in 2D) allowing for stable lattice structures. This contrasts with the $Z_{2}$ model where the dynamics are typically dominated by the evolution of an single large domain with closed walls inside it. In earlier work we have shown that regular lattices can be constructed which are elastically stable if a wall network supports junctions and have begun to assess the cosmological implications of such a network~\cite{BCCM,BM1,BCM,BM2}. This work was based on a continuum approximation representing the lattice macroscopically as an elastic medium, analogous in many ways to soft condensed matter systems such as foam and soap films. Here, we are concerned with whether the formation of a regular lattice can occur from random initial conditions in a full field theory and whether the total number  of the walls respects the same $t^{-1}$ scaling behaviour of the $Z_{2}$ model, or whether they can frustrate and are dragged along with the Hubble flow and the number of walls scales like $a^{-1}$.  

The problem with numerical simulations of defects in an expanding universe is that there are the two very different length scales involved - the defect thickness and separation scale. The ratio of the former to latter rapidly becomes too small to simulate accurately and so the common procedure taken is to fix the defect size in co-moving coordinates whilst adjusting the equation of motion to ensure momentum conservation~\cite{RPS}. It has been argued that this procedure does not significantly effect the evolution. We choose to evolve our simulations in Minkowski space to avoid making this approximation which should be accurate so long as the conclusions we make are based on epochs before the light crossing time. If a system scales in Minkowski spacetime, we would expect it to do so in an expanding universe.

The paper is organised as follows. In section~\ref{cam} we introduce the cubic anisotropy model and review the features of the model relevant for domain wall formation. Section~\ref{num} then contains details of the numerical simulations and how we extract the results. In section~\ref{rlc} we show how simple regular lattices can be constructed and investigate their stability with respect to the continuum approximation. We then present results for the dynamical evolution of walls in the $Z_{2}$ and cubic anisotropy models in sections~\ref{z2} and~\ref{camres} respectively. Finally, we discuss the implications of our results and conclude.    

\section{Cubic anisotropy model} \label{cam}

We will consider models for a global vector field $\phi=(\phi_1,..,\phi_N)$ with standard relativistic kinetic term and potential
\begin{equation} \label{cameqn}
V(\phi)={\lambda\over 4}\left(|\phi|^2-\eta^2\right)^2+\epsilon\sum_{i=1}^{N}\phi^4_i\,.
\end{equation}
If $\epsilon=0$ then the model has $O(N)$ symmetry, but this is broken to a discrete (hyper)cubic symmetry when $\epsilon\ne 0$. Critical points of the potential satisfy 
\begin{equation} \label{critpoints}
\lambda\phi_i\left(|\phi|^2-\eta^2\right)+4\epsilon\phi^3_i=0\,,
\end{equation}
for each $i=1,N$. The case of $N=1$ and $\epsilon=0$ corresponds to the standard case of $Z_{2}$ symmetry breaking. In the case of $N=2$, the potentials for $\epsilon>0$ and $\epsilon<0$ are, from the point of view of the structure of the vacuum manifold, essentially the same modulo a rotation. If we write the field as $(\phi_1,\phi_2)=|\phi|(\cos\theta,\sin\theta)$ then, in the case where $\epsilon>0$, there are four minima at $\theta=\{\pi/4, 3\pi/4, 5\pi/4, 7\pi/4 \}$, whereas for $\epsilon<0$ the minima are at $\theta=\{0, \pi/2, \pi, 3\pi/2 \}$. In fact the potential can be written as 
\begin{equation} \label{axion}
V(|\phi|,\theta)={\lambda\over 4}\left(|\phi|^2-\eta^2\right)^2+{\epsilon \over 4}|\phi|^4\left(3+\cos 4\theta\right)\,, 
\end{equation} 
and hence for this special case the model is similar to an axion domain wall model~\cite{S,VE}.

\begin{figure}[] 
\centering
\mbox{\resizebox{0.3\textwidth}{!}{\includegraphics{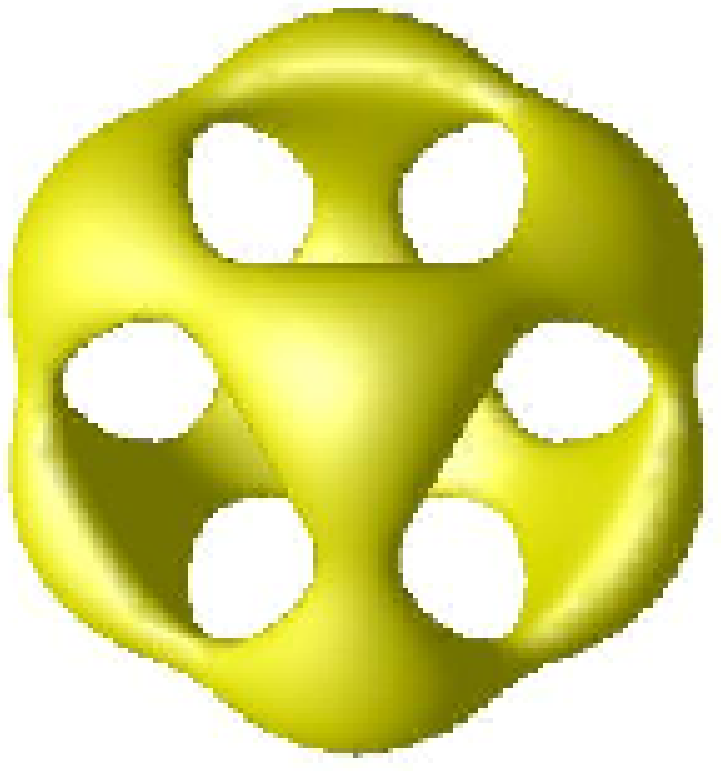}}}
\mbox{\resizebox{0.3\textwidth}{!}{\includegraphics{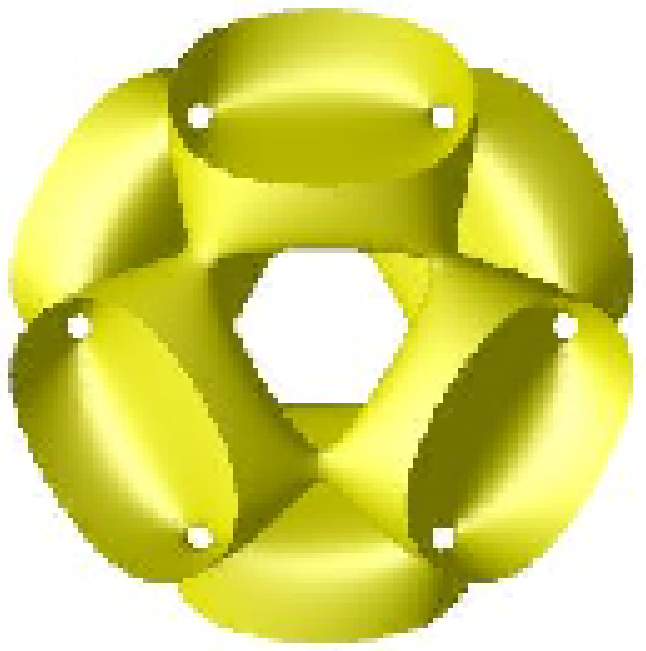}}}
\caption{Structure of the vacuum manifold from a broken O(3) symmetry illustrated by a potential isosurface. The isosurface on the left has an anisotropy parameter $\epsilon=0.08$ and the minima reside in the vertices of the cube. The isosurface on the right has $\epsilon=-0.08$ and the minima reside in the faces of the cube (or the vertices of an octahedron).}
\label{cube}
\end{figure}

When $N\ge 3$ the structure of the vacuum manifold is different in the two cases $\epsilon>0$ and $\epsilon<0$ . This is illustrated in Fig.~\ref{cube} for $N=3$. In the $\epsilon>0$ case the minima reside at the vertices of a (hyper)cube and in the $\epsilon<0$ case the minima reside at the faces (or the vertices of an (hyper)octahedron - the dual of the cube). A variety of different types of wall system are then possible. For the broken $O(3)$ model with $\epsilon>0$ a total of three different types of walls can be produced, each with different tension. The lightest walls correspond to interpolation across an edge, such as $\{+,+,+\} \rightarrow \{+,+,-\}$. Heavier walls correspond to interpolation across a face, for example $\{+,+,+\} \rightarrow \{+,-,-\}$, and across the body diagonal, for example $\{+,+,+\} \rightarrow \{-,-,-\}$. For general $N$ there are $2^{N}$ discrete minima and $N$ different types of walls are possible. In the broken $O(3)$ case with $\epsilon<0$ two types of walls can be produced with different tension. The lightest walls correspond to interpolation between faces in different perpendicular directions, such as $\{+,0,0\} \rightarrow \{0,+,0\}$.  Heavier walls correspond to interpolation between faces in the same direction, for example $\{+,0,0\} \rightarrow \{-,0,0\}$.  For general $N$ there are $2N$ discrete minima, but only two different types of wall are possible.

\begin{figure}[] 
\centering
\mbox{\resizebox{1.0\textwidth}{!}{\includegraphics{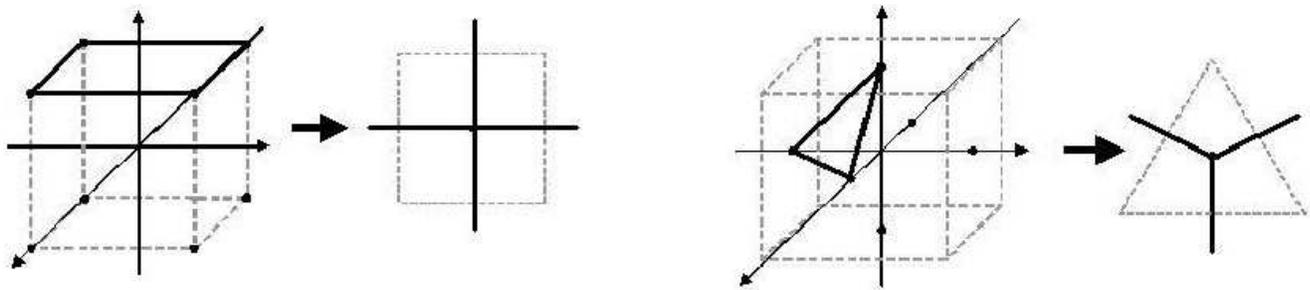}}}
\caption{Wall-string junctions in the $N=3$ cubic anisotropy model with $\epsilon>0$ (left) and $\epsilon<0$ (right). The diagrams show how the arrangement of fields on the vacuum manifold gives rise to wall junctions. In the $\epsilon>0$ model fields can arrange themselves, for example, on one of the faces of the vacuum manifold. We illustrate how this arrangement gives rise to X type junctions with four walls meeting at a string. In the $\epsilon<0$ model Y type junctions can arise with three walls meeting at a string/vortex by connecting three faces as illustrated.}
\label{walltypes}
\end{figure}

The structure of the vacuum manifold means that there is an important distinction in the types of walls produced in the $\epsilon>0$ and $\epsilon<0$ models. In the former, X type junctions will form (such as four walls meeting at a string/vortex) whilst in the latter, Y type junctions will form (such as three walls meeting at a string/vortex). These configurations are illustrated in Fig.~\ref{walltypes}. In previous work attention has been drawn to the difference in the stability properties of these systems~\cite{BCCM,BCM}. In the Y type junction there is no freedom of adjustment in the equilibrium angle of intersection at the string which must be $2\pi/3$ if the wall tensions are equal. In the X type junction equilibrium is maintained by opposing pairs with the same tension, even if the angle of intersection is not equal to $\pi/2$. It was pointed out that X type junctions are more favourable for the stability of a lattice. We note that claim of ref.~\cite{avelino1} that if all the walls have the same tension, then only Y type junctions are possible does not apply here.

Models with $\epsilon<0$, giving rise to Y type junctions, have been previously studied in ref.~\cite{K}. These simulations suggest some evidence of the formation of cellular hexagonal structures and it was suggested that there are deviations from the standard self-similar scaling. However, these deviations are only slight and only a small number of simulations with limited dynamical range were considered.

\section{Numerical methodology} \label{num}

\subsection{Discretisation}

The equations of motion were computed by minimising the action
\begin{equation}
S=\int d^{4}x \sqrt{-g} \left( \frac{1}{2} (\partial_{\mu} \phi_{i})^{2} - V(\phi_{i}) \right)\,,
\end{equation}
where the potential energy is given by~(\ref{cameqn}). For the Minkowski metric $\eta_{\mu \nu}= {\rm diag}(-1,1,1,1)$, we obtain the Euler-Lagrange equation
\begin{equation} \label{euler}
\ddot{\phi}_{i}-\nabla^{2} \phi_{i} + \lambda\phi_i\left(|\phi|^2-\eta^2\right)+4\epsilon\phi^3_i=0.
\end{equation}
This equation has been discretized on a regular, two dimensional cubic grid with $P$ points in each direction and a spatial separation size $\Delta x$. The simulations are evolved in two dimensions since this allows a larger dynamical range. Furthermore, we expect the scaling behaviour in two dimensions to represent an upper limit on the formation of stable domains in three dimensions due to the increased dynamical freedom for the decay of walls in three dimensions. We evolve~(\ref{euler}) using a second order time, fourth order space (2-4) leapfrog algorithm to accurately represent any large spatial gradients associated with the domain walls. For further details of this particular algorithm, see ref.~\cite{BSut}. The boundary conditions we employ are periodic, and so in obtaining any quantitative information regarding the evolution of the wall network we limit the total time of the simulation to the light crossing time. This is the time taken for two signals emitted from the same point travelling in opposite directions to interfere with each other, and is given in terms of the number of simulation timesteps $T$ by $P \Delta x / (2 \Delta t)$. The simulations in this paper typically use $P=2048$ (making them some of the largest reported to date) with $\Delta x=0.5$ and $\Delta t=0.1$, which means that the light crossing time is $T \sim 5000$. We use larger grids and vary grid resolution parameters where necessary to test the veracity of results. The choice of $P$, $\Delta x$ and $\Delta t$ are important for a number of reasons. The first is that the 2-4 algorithm must be stable and well posed. The standard Courant condition for a linear equation in two dimensions requires that $\sqrt{2} \Delta t < \Delta x$. The nature of~(\ref{euler}) leads to a modification of this condition and we choose $\Delta t$ for a given $\Delta x$ by trial and error, using the Courant condition as an upper bound. Secondly, one must consider the wall thickness $\delta$. The wall cannot be too thin, otherwise the discretisation will not be able to fully resolve the wall profile, which  places an upper limit on $\Delta x$. On the other hand, the wall cannot be too thick as the typical wall separation scale $l$ must exceed $\delta$. We found a suitable full-width-half-maximum (FWHM) of the wall profile was typically five grid points.

\subsection{Initial Conditions}\label{initial}

Initial conditions were created by two methods. Both involve populating each grid point with a value of $\phi_{i}$ and $\dot{\phi}_{i}$ according to some prescription. In section~\ref{rlc} we show how regular tilings of domain walls can be constructed and so suitable initial conditions can be generated using a Voronoi tessellation of the plane (sometimes known as a Dirichlet tessellation). This partitions the plane into convex polygons such that each polygon contains a single generating vacua and each subsequent vacua in a given polygon is closer to its generating vacua than any other.  The Voronoi tessellation of a set of triangular points, for example, gives a regular hexagonal lattice. 

The second method we use is a random distribution of $\phi_{i}$ with a given correlation length, which is more realistic from a cosmological viewpoint. This was achieved by distributing all possible vacua in a given model in cubic domains covering $c$ grid squares. Since this correlation size introduces an unwanted length scale into the problem we vary this parameter according to $c=1,2,4,8$ to check for any dependence of the late time dynamics on initial conditions. The initial value of $\phi_{i}$ is randomly chosen to be one of the potential minima specified by~(\ref{critpoints}) and the initial value of $\dot{\phi}_{i}$ was set to zero. Since our main objective is to observe the late time dynamics the only requirement of the initial conditions is to distribute the value of $\phi_{i}$ randomly. In this way the typical variance of the particular minima which each mesh point inherits is less than $0.1\%$ over the entire grid. Any particular bias towards one vacua can significantly affect the late time dynamics of the network~\cite{CLO,LSW}.

\begin{figure}[] 
\centering
\mbox{\resizebox{0.45\textwidth}{!}{\includegraphics{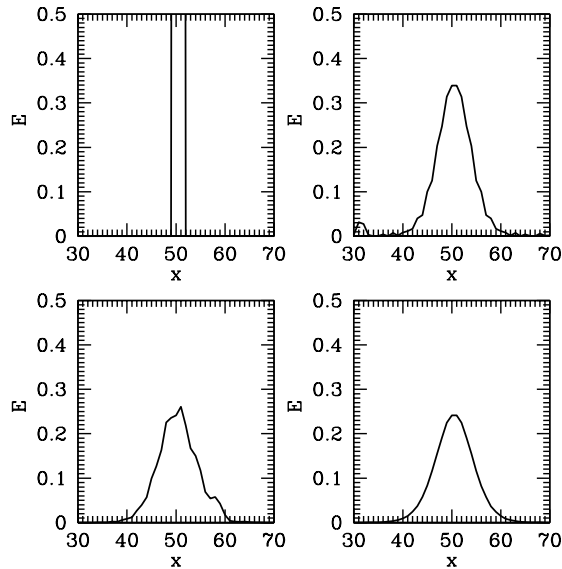}}}
\caption{Energy density through a static wall in the $Z_{2}$ model for increasing timesteps $T$ = 0 (top-left), 60 (top-right), 120 (bottom-left), 180 (bottom-right). The initial spike in energy is due to the initial sharp discontinuity between vacua over neighbouring grid points.}
\label{stepslice}
\end{figure}

At the beginning of the evolution~(\ref{euler}) is modified to include a dissipative term. This dissipation smooths out the initial sharp discontinuity between vacua over neighbouring grid points without creating unwanted radiation which can confuse the interpretation of the late time dynamics. Fig.~\ref{stepslice} shows, for example, the smoothing out of a an initially sharp static domain wall in the $Z_{2}$ model under dissipative dynamics. The level of dissipation is set at 0.5 and after approximately 200 timesteps the wall profile has relaxed. We then remove this dissipation over the next 50 timesteps in a linear fashion and then let the system freely evolve. In this way this dissipative regime lasts for approximately $5 \%$ of the total simulation time. We have checked for smoothing of walls for $N \ge 1$ and again an initial dissipative period of approximately 200 timesteps is sufficient. From measurements of the wall thickness we find that when $N \ge 1$ the wall thickness $\delta \approx |\epsilon|^{-1/2} \, \eta^{-1}$ when  $\epsilon < 0$ and $\delta \approx (N/ \epsilon)^{1/2} \, \eta^{-1}$ when $\epsilon > 0$.

\subsection{Evaluation of number of walls}

The total number of the walls in the  network is fitted to the power law $n \propto t^{-\alpha}$ (in the analysis we fit to the number of simulation timesteps $T$ since $t= T \Delta t$) and the density of walls is proportional to this, assuming Nambu-Goto walls. In order to compute the scaling exponent $\alpha$ we have computed the number of each variety of wall in a given model. This was achieved by the following method. At each lattice site we take the value of $\phi_{i}$ and compute which minima on the vacuum manifold the field is nearest to. We then compute the same quantity on an adjacent lattice site. If the nearest minima is the same on adjacent lattice sites this does not constitute a wall. If the nearest minima is different then we assume there is a wall and increment the wall count by one. In the case of the $Z_{2}$ model this simply gives an estimate of the surface on which $\phi=0$. When $N \ge 2$ there are walls with different tensions and we are able to keep track of the scaling properties of each wall type. This method of computing the number of walls was tested by using regular tilings of domain walls with a known surface area.

The discrete method of computing the number of walls leads to an overestimation of the area compared to the continuous value, as one is approximating a smooth curve by a set of line segments. A simple estimate of this factor can be computed by considering a straight wall link at some orientation in space. The estimation error is dependent on the orientation and so an average factor can be obtained by integrating over all possible orientations of the wall link. In two dimensions this overestimation factor is found to be $4/\pi$. Since our main concern is computing the scaling exponent $\alpha$, and not the scaling density, this factor is negligible if there are a large number of walls in the box to sample from.

\subsection{Scaling Exponent Evaluation}

In order to compute the scaling exponent one needs to identify a range over which to fit the data. One expects a transient regime at the start of the simulation  as the system relaxes after the removal of the dissipative term. On the other hand, toward the end of the simulation finite size effects may become important. We therefore expect the optimum fit to occur near the mid-point of the simulation. For completeness when quoting results, however, we compute the exponent in bins of time and also vary the size of these bins. For each simulation we perform ten runs and compute a mean and variance of the scaling exponent using a least squares regression of data on the number of walls.

\section{Regular lattice configurations} \label{rlc}

\begin{figure}[] 
\centering
\mbox{\resizebox{0.5\textwidth}{!}{\includegraphics{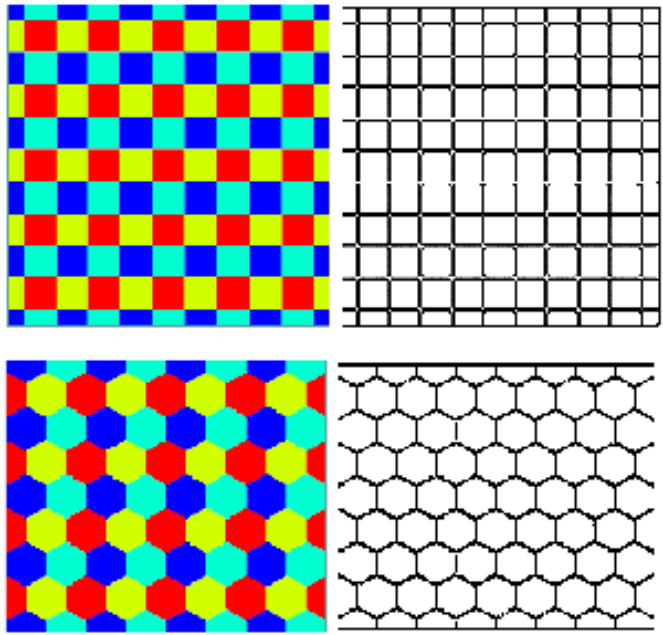}}}
\caption{Regular tilings of the plane constructed from the cubic anisotropy model. The square tilings on the top are constructed from a broken O(2) model with $\epsilon>0$ and the hexagonal tilings on the bottom from a broken O(4) model with $\epsilon<0$. The figures on the left show distinct vacuum states in different colours/shades of grey and the figures on the right the correspondingly energy density associated with each wall.}
\label{lattice}
\end{figure}

The ability of the vacuum manifold to support junctions means that regular tilings of the plane can be constructed and it is interesting to study their stability. In the case of $\epsilon>0$, X type junctions can be used to construct a square type lattice, while the Y type junctions in the $\epsilon<0$ model can be used to construct a hexagonal lattice. These are illustrated in Fig.~\ref{lattice}, which shows the output of numerically solving the field equations. Parameter values were chosen so that the characteristic wall thickness $\delta$ was far smaller than the separation scale $l$ of the lattice. Initial conditions to correspond to the relevant tiling configuration, set up using the Voronoi tesselation method, and an initial period of dissipation was applied to allow the walls to relax from the initial discontinuity. The system was then left to freely evolve and the output in Fig.~\ref{lattice} shows the network after a time period of several light crossing times. In each case a different colour represents a distinct vacuum state. In subsequent sections, where we compute scaling exponents the simulations are terminated after a single light crossing time, but for the purpose of demonstrating stability it is acceptable to run the simulations for longer. Indeed, running the simulations for longer can only make clearer the stability properties. For $\epsilon>0$ the minimal model required to tile the plane with a square lattice is $O(2)$, corresponding to four domain walls with equal tension and a single type of string. For $\epsilon<0$ the minimal model required to tile the plane with a hexagonal lattice is $O(4)$, corresponding to four domain walls with equal tension and two types of strings, again with equal tension. The necessity to use $O(4)$ arises from the use of periodic boundary conditions in the numerical simulations. In the unperturbed case with exact square or hexagonal symmetry both lattices are stable.   

In previous work the macroscopic elastic properties of both the hexagonal and square lattices were discussed~\cite{BCM}. The relevant rigidity coefficients for each lattice were computed and these were used to evaluate the propagation velocities for macroscopic scale perturbation modes. In both cases a marginally stable zero mode was shown to exist. In the case of the square lattice this zero mode occurs for a finite number of directions and was shown to be of infinite extent and not sufficient for the construction of locally confined perturbations which might lead to an instability. In particular, it was shown that the zero mode corresponds to the walls moving toward each other at a constant velocity. However, in the case of the hexagonal lattice the zero mode exists for all wavenumber directions and can provide a sufficient basis for locally destabilising perturbations. 

\begin{figure}[] 
\centering
\mbox{\resizebox{0.8\textwidth}{!}{\includegraphics{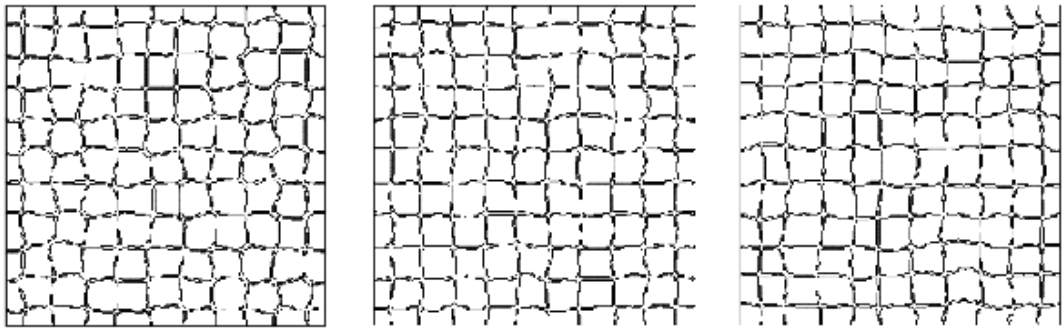}}}
\caption{Energy density of a perturbed square lattice at subsequent time slices of the simulation. The figures on the left, middle and right shows the configuration at 0, 2 and 4 light crossing times respectively.}
\label{cubepert}
\end{figure}

\begin{figure}[] 
\centering
\mbox{\resizebox{0.8\textwidth}{!}{\includegraphics{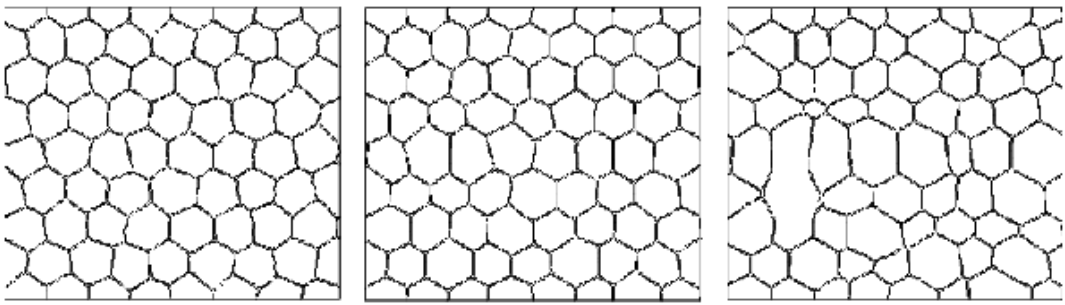}}}
\caption{Energy density of a perturbed hexagonal lattice at the same time slices as for the perturbed square lattice.}
\label{hexpert}
\end{figure}

These analytic results can be tested in the context of the cubic anisotropy model studied here. Each lattice was perturbed from its initial tiling configuration and after an initial period of dissipation was left to evolve freely. The energy density of each network at various time slices in the simulation are shown in Figs.~\ref{cubepert} and~\ref{hexpert}. The overall structure of the square lattice remains intact at the end of the simulation. The zero mode is apparent as there is nothing preventing parallel walls drifting together. If the walls did drift too far and annihilate each other the consequent annihilation would, however, not destabilise the entire lattice. For the hexagonal lattice the perturbation has rather more serious consequences. The zero mode allows hexagons to reduce in size with no change in energy. When the domain is sufficiently small in size, there is annihilation between walls. Running the simulation beyond the last frame shown in Fig.6 confirms that the endpoint is a single vacuum state. This local mode then continues to grow and subsequently destroys the entire lattice. Both of these conclusions appear to support the basic picture presented in ref.~\cite{BCM}.

From a cosmological point of view, the tuning required to construct stable lattices as we have done here is a severe limitation of the model. However, studying the properties of lattice configurations can help in the interpretation of results when evolving the equations of motion from random initial conditions.

\section{Standard case of $Z_2$ symmetry breaking} \label{z2}

\begin{figure}[] 
\centering
\mbox{\resizebox{0.8\textwidth}{!}{\includegraphics{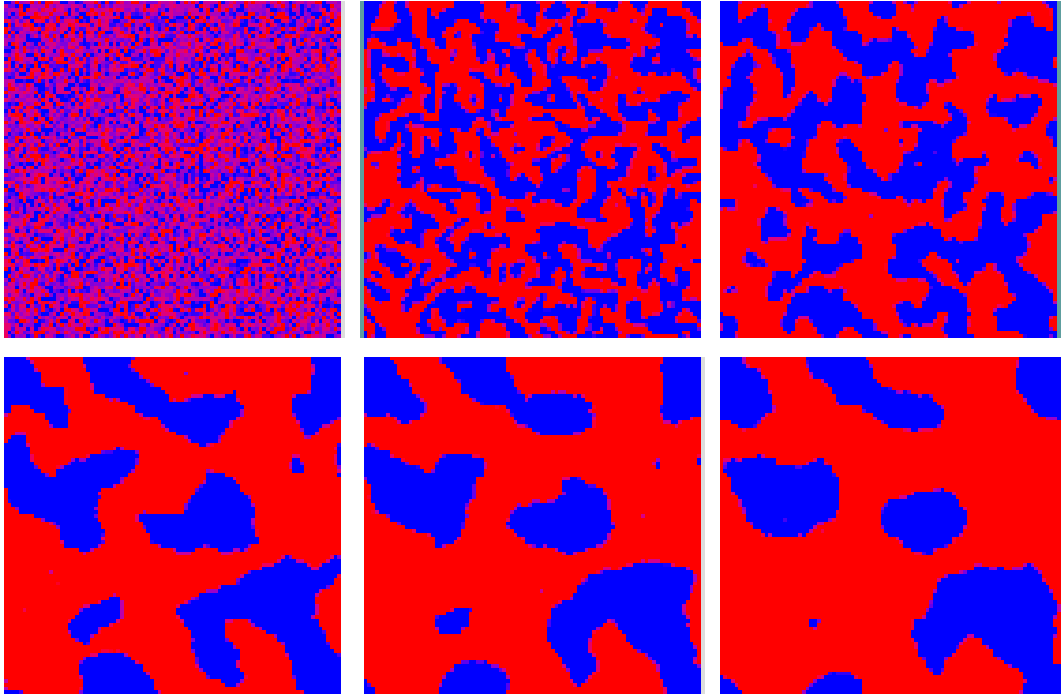}}}
\caption{Distribution of vacua in the $Z_2$ model. The time slices are taken in steps of 500 starting from the initial state (top-left to bottom-right). The different colours/shades of grey show which of the two minima on the vacuum manifold the field is nearest to. Blue/dark-grey regions represent the minima \{-\} and red/light-grey regions the minima \{+\}. The length of the boundary between colours gives an estimate of the number of walls.}
\label{o1time}
\end{figure}

In order to calibrate and validate our results for $N \ge 2$ (presented in the next section) we have performed an extensive suite of simulations in the $N=1$, $\epsilon=0$ case. The simulations use a grid size of $2048^{2}$, $\Delta x=0.5$ and $\Delta t=0.1$ with the potential parameters $\lambda=\eta=1$ which corresponds to the general case by an overall scaling of the energy and length of walls. The FWHM wall profile in this particular case is then approximately four grid points. We populate the initial grid with random initial conditions using the second method described in section.~\ref{initial} and the dynamical evolution for a sample simulation is shown in Fig.~\ref{o1time}. The formation of an infinite domain is clearly apparent containing what are effectively closed walls, with a single vacuum state eventually occupying the entire grid. In table~\ref{tab:z2corr} we present results for the scaling exponent $\alpha$ as a function of the initial correlation size. The scaling exponent is computed in bins of 500 timesteps, where the time quoted is the central value in each bin. Results are computed from an ensemble of ten simulations with different seeds for the initial conditions. To gain some idea of the evolution of the wall network, the results in each bin can be compared with the output for a single run at the equivalent timestep in Fig.~\ref{o1time}. 

\begin{figure}[] 
\centering
\mbox{\resizebox{0.35\textwidth}{!}{\includegraphics{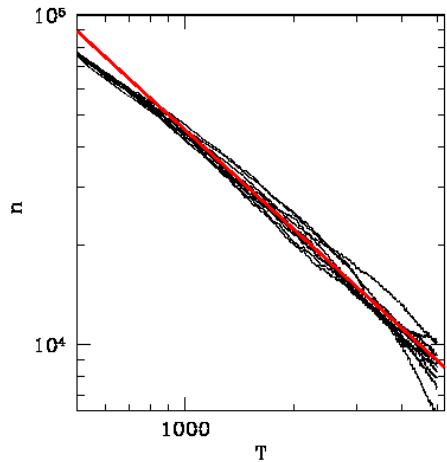}}}
\caption{Evolution of the number of walls $n$ as a function of simulation time $T$ in the $Z_{2}$ model for each of the ten simulations with $c=2$. Also plotted for comparison is $t^{-1}$ scaling (thick red line).}
\label{noz2walls}
\end{figure}

Results are in broad agreement with other work~\cite{RPS,CLO,LSW,GH,AM1,AM2}, validating our methods. In particular, there are several common features which warrant some discussion. The most significant is evolution of the scaling exponent. There has been some suggestion of a logarithmic correction to this exponent~\cite{RPS}. We find hints of some evolution of the scaling exponent but within the error limits the results are inconclusive. Indeed, even with very large simulations it is difficult to pinpoint any such correction~\cite{AM1}. The approach to scaling is relatively slow, requiring several hundred timesteps after the termination of the dissipative regime. This is illustrated in Fig.~\ref{noz2walls}, where we plot each of the ten runs for $c=2$. The results are consistent with $t^{-1}$ scaling after approximately 1000 timesteps. Referring to table~\ref{tab:z2corr}, evaluation of the scaling exponent in the first time bin gives a value consistent with the results of~\cite{RPS,CLO} and the following two bins a value more compatible with~\cite{LSW,GH,AM1,AM2}. The error bars then start to increase significantly as a single domain begins to dominate the simulation. The similarity of the scaling exponents obtained for different values $c$ gives confidence that any residual `memory' of the initial length scale is erased by the time we begin regression. Indeed, the similarity of the scaling exponents for different initial conditions provides some evidence that the scaling is self-similar and any deviation from this is due to limited dynamical range. In order to compare results with those when $N \ge 1$ we use the combined results, for different values $c$, listed in table~\ref{tab:z2corr}. Rather than choose a single absolute value for the scaling exponent we compare results over time bins. 

\begin{table}[] 
\begin{center}
\begin{tabular}{|c|c|c|c|c|} \hline
& \multicolumn{4}{|c|}{$T$} \\ \hline 
\hspace{0.3cm} c \hspace{0.3cm} &  1000 &  1500 &  2000 &  2500  \\ \hline

 1 & $0.86 \pm 0.06$ & $0.98 \pm 0.10$ & $0.97 \pm 0.12$ & $1.14 \pm 0.17$ \\ \hline

 2 & $0.87 \pm 0.07$ & $0.96 \pm 0.06$ & $0.97 \pm 0.07$ & $0.99 \pm 0.21$ \\ \hline

 4 & $0.88 \pm 0.07$ & $0.96 \pm 0.06$ & $0.97 \pm 0.07$ & $0.99 \pm 0.22$ \\ \hline

 8 & $0.87 \pm 0.06$ & $0.95 \pm 0.10$ & $0.98 \pm 0.09$ & $1.04 \pm 0.19$ \\ \hline \hline

 ALL & $0.87 \pm 0.06$ & $0.96 \pm 0.08$ & $0.97 \pm 0.09$ & $1.04 \pm 0.20$ \\ \hline 

\end{tabular}
\end{center}
\caption{\label{tab:z2corr} Mean and $1 \sigma$ values of the scaling exponent $\alpha$ in different time bins for variable initial correlation sizes.}
\end{table}

\section{Simulations of the cubic anisotropy model} \label{camres}

\begin{figure}[] 
\centering
\mbox{\resizebox{1.0\textwidth}{!}{\includegraphics{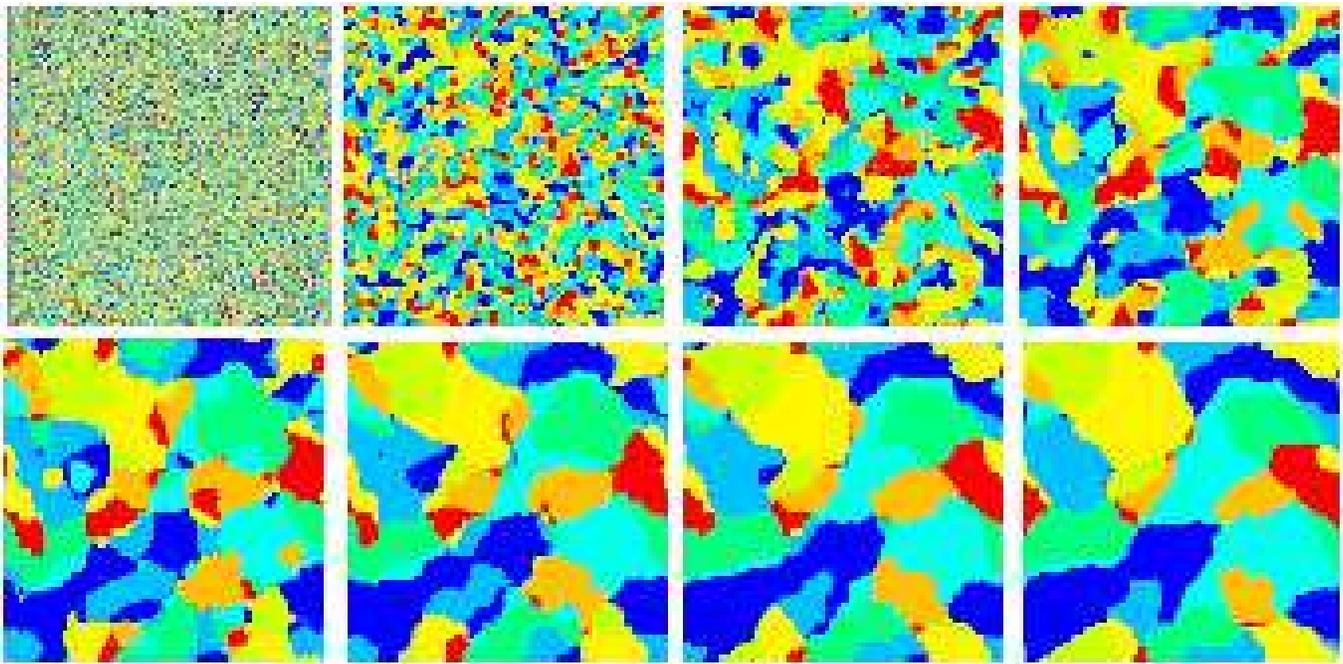}}}
\caption{Distribution of vacua in the $N=3$ model with $\epsilon=0.1$. The time slices are taken in steps of 500 starting from the initial state (top-left to bottom-right). The different colours/shades of grey show which of the eight minima on the vacuum manifold which the field is nearest to.}
\label{o3_bc_time}
\end{figure}

\begin{figure}[] 
\centering
\mbox{\resizebox{0.35\textwidth}{!}{\includegraphics{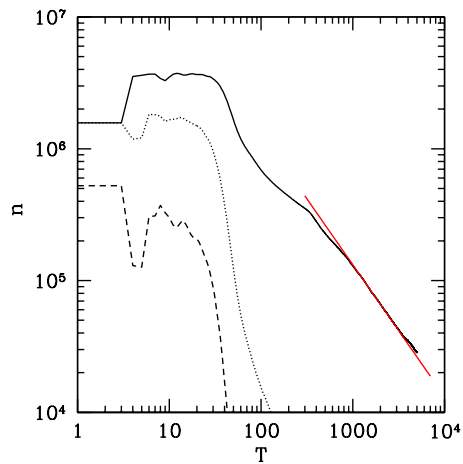}}}
\caption{Evolution of the number of walls in the $N=3$ model with $\epsilon=0.1$. There are three types of wall in this model - walls of low tension (solid-black line) , intermediate tension (dotted) and high tension (dashed). The low tension walls enter a scaling regime while the walls of high tension decay off rapidly. Also plotted for comparison is $t^{-1}$ scaling (thick red line).}
\label{o3_walls}
\end{figure}

We now turn our attention to the case where $N \ge 2$. The wall evolution is quantitatively different to the $N=1$ model where the dynamics of the network become dominated by the evolution of a single domain. Here, the existence of non-trivial junctions modifies the dynamical behaviour. Simulation parameters are chosen as in the $N=1$ case and Fig.~\ref{o3_bc_time} shows the evolution of an $N=3$ model with $\epsilon=0.1$ using an initial correlation size of two grid points ($c=2$). The initial network is very dense with a large number of domain walls. The density of walls is greater than in the $N=1$ model at an equivalent timestep. This is due to the vacuum manifold containing $2^{3}$ discrete points admitting a variety of domain wall configurations rather than just a single wall as in the $N=1$ case. Since three different types of wall can be produced in the $N=3$ model, a simple counting exercise reveals that in a random distribution $3/7$ of the initial wall density will correspond to walls of the lowest tension, $3/7$ will have intermediate tension and $1/7$ the highest tension. This is confirmed in Fig.~\ref{o3_walls}, where we plot the total number of walls for each variety as a function of time for the single simulation depicted in Fig.~\ref{o3_bc_time}. There is an initial conversion of high tension walls to low tension walls - after only a small number of timesteps the number of low tension walls has increased by a factor of two. The number of high tension walls decays very rapidly with the system attempting to minimise energy by rearranging the distribution of vacua on the vacuum manifold. 

A favourable configuration is four low tension walls meeting at a string, corresponding to the vacua arranging themselves on a face of a cube on the vacuum manifold.  The appearance of such configurations can be clearly seen in Fig.~\ref{o3_bc_time}. In these X type junctions equilibrium is maintained by opposing walls of the same tension. However, the network shows no sign of the formation of stable lattices - the walls continue to contract under their own surface tension and decay processes are efficient. As noted in section~\ref{rlc}, there is nothing to prevent parallel walls from drifting together and annihilating each other. Referring to Fig.~\ref{o3_walls}, after some initial transient, the low tension walls enter a scaling regime as they undergo a self-similar rearrangement process. 

\begin{figure}[] 
\centering
\mbox{\resizebox{1.0\textwidth}{!}{\includegraphics{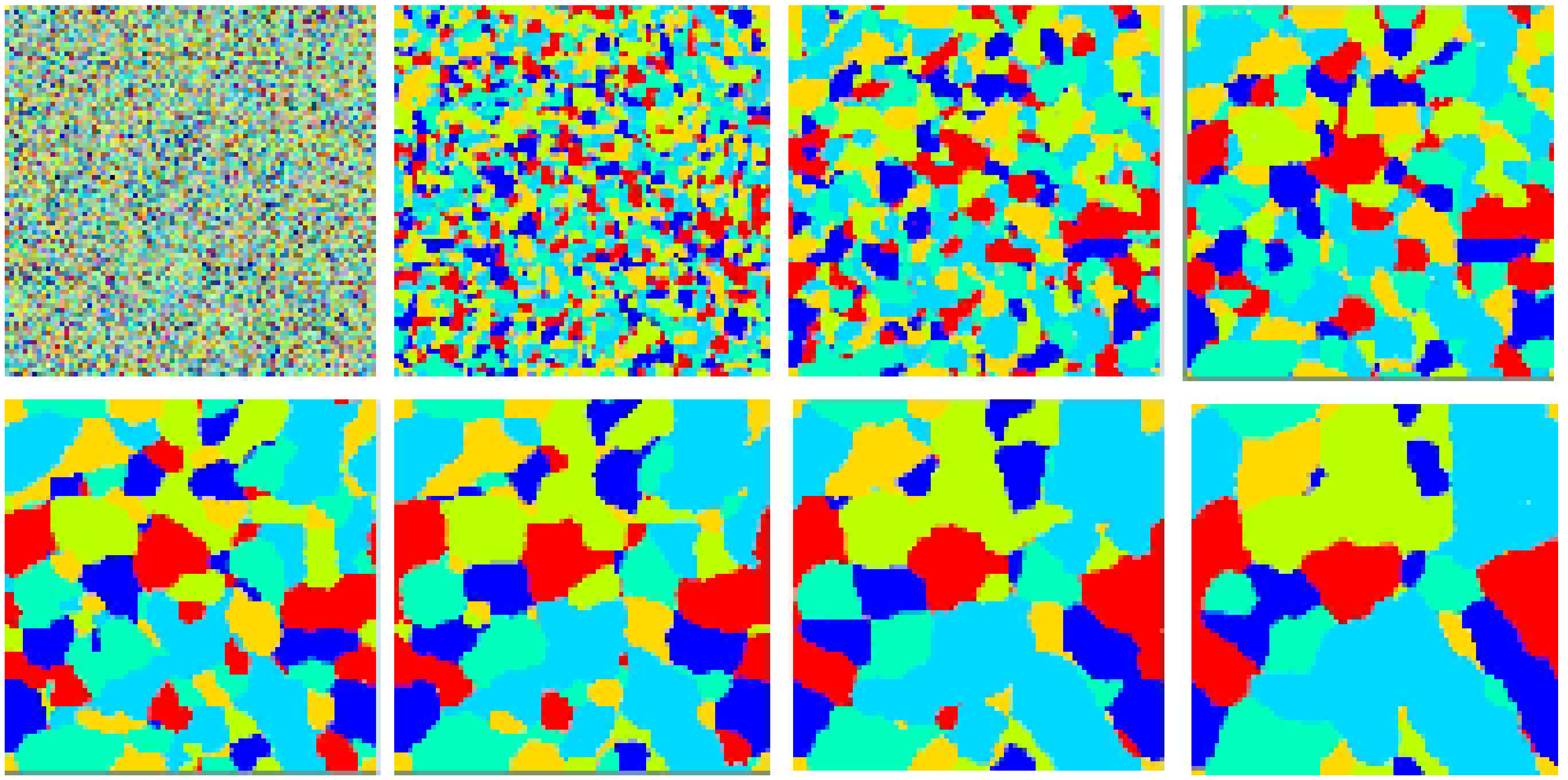}}}
\caption{Distribution of vacua in the $N=3$ model with $\epsilon=-0.1$. The time slices are taken in steps of 500 starting from the initial state (top-left to bottom-right).}
\label{o3_kc_time}
\end{figure}

Fig.~\ref{o3_kc_time} shows the evolution of an $N=3$ simulation with $\epsilon=-0.1$. The initial network is not as dense as the $\epsilon=0.1$ case due to a smaller number of minima on the vacuum manifold.  Two types of wall can be produced when $\epsilon<0$ and in the $N=3$ model 5/6 of the initial wall density corresponds to walls of the lowest tension. The high tension walls again decay rapidly into low tension walls over a small number of timesteps. A favourable configuration in this case is three light walls meeting at a string at an intersection angle of $2\pi/3$, corresponding to a triangular arrangement on the vacuum manifold and a hexagonal wall lattice. However, the dynamics of the network is again dominated by the contraction of walls under their surface tension contrary to the suggestions made in ref.~\cite{K}. No regular hexagonal domains form and domains frequently join as the network coarsens.

In order to quantify whether the scaling relation is the same as the $N=1$ model we perform ten runs for each set of parameters and have computed the average scaling exponent. We investigate models with $2\le N\le 7$ to check for any dependence on the number of vacua. It is possible, for example, that the increased number of vacua for large $N$ will lead to a different annihilation probability between domains, which could slow down the rate of domain growth. We consider the symmetry breaking parameter $\epsilon = \pm 0.1$ which gives rise to the two different types of network - those which form only X type junctions $(\epsilon>0$) and those which form Y type junctions ($\epsilon<0$). The late time dynamics were again found to be independent of the initial conditions and so for clarity we present average scaling exponents. These results are presented in table~\ref{tab:on} and are summarised below:

\begin{itemize}
\item{Typically, in most of the models $\alpha < 1$, although not by much. These deviations are not statistically significant in individual time bins, but it is interesting to note the general trend for $\alpha < 1$. This is a feature of most measurements of the scaling exponents reported to date, which is most likely due to finite size effects. In the $N=1$ model we found hints of a weakly increasing scaling exponent as a function of time. Here, in some cases, there are hints of a weakly decreasing scaling exponent, but the limitation of statistical errors makes it difficult to give any firm conclusions.}
\item{In the $\epsilon=-0.1$ models there is no significant change in $\alpha$ as a function of $N$, in conflict with the results of ref.~\cite{K}. The larger number of vacua does not reduce the annihilation probability between adjacent domains such that the coarsening of the network is any slower. The network appears to enter the scaling regime quickly - at 1000 timesteps with $\alpha \approx 1$ and small error bars. There is an increase in these error bars as a function of time as in the $N=1$ model. In the $N=1$ model these increased error bars were  attributed to a single vacua dominating the box size. In the models studied here there are still a large number of walls present in the box to sample from and so a sample variance error in measuring the number of walls seems less likely. The ability of junctions to form domains means that the variance between simulations could be due to the network actually physically scaling differently depending on the particular arrangement of domains. Given the range of values for the scaling exponent which we found for the $Z_2$ model, which is believed to scale,  we conclude that these models are very much compatible with scaling.}
\item{For $\epsilon=0.1$ the situation is less clear. The values computed for $\alpha$ are on the whole lower, particularly for larger values of $N$, although they are still not significantly less than 1 for all times. We also observe that for increasing $N$ there is a longer intial transient regime. This effect appears to weaken with increasing time and might be attributed to the thickness of the walls. In the case where $\epsilon<0$ the wall thickness does not vary with $N$ - the value of $\epsilon=-0.1$ used here gives rise to a FWHM wall profile of approximately five grid points. In the case where $\epsilon>0$ the wall thickness scales approximately as $N^{1/2}$ - in the case where $N=5$, for example,  the FWHM wall profile increases to approximately 11 grid points when $\epsilon=0.1$ and the ratio of the wall separation to thickness is decreased. In more physically realistic situations, the thickness of the walls should not affect their dynamics when well separated. }
\end{itemize}

\begin{table}[]
\begin{center}
\begin{tabular}{|c|c|c|c|c|c|c|c|c|} \hline
 \multicolumn{2}{|c|}{} & \multicolumn{7}{|c|}{Time} \\ \hline 
\hspace{0.3cm} $N$ \hspace{0.3cm} &  $\epsilon$ & 1000 &  1500 &  2000 &  2500 & 3000 & 3500 & 4000 \\ \hline \hline

 2 & 0.1  & $0.92 \pm 0.04$ & $1.00 \pm 0.09$ & $0.99 \pm 0.09$ & $0.95 \pm 0.14$ & $0.97 \pm 0.16$ & $1.02 \pm 0.17$ &  $1.00 \pm 0.21$ \\ \hline

 2 & -0.1  & $0.91 \pm 0.04$ & $0.96 \pm 0.08$ & $0.98 \pm 0.12$ & $0.94 \pm 0.13$ & $1.05 \pm 0.16$ & $1.05 \pm 0.28$ &  $1.09 \pm 0.25$ \\ \hline \hline

 3 & 0.1  & $0.88 \pm 0.03$ & $0.99 \pm 0.06$ & $0.99 \pm 0.05$ & $0.93 \pm 0.14$ & $0.90 \pm 0.10$ & $0.91 \pm 0.15$ &  $0.90 \pm 0.17$ \\ \hline

 3 & -0.1  & $0.98 \pm 0.04$ & $0.98 \pm 0.07$ & $0.93 \pm 0.09$ & $0.95 \pm 0.11$ & $0.96 \pm 0.15$ & $0.98 \pm 0.17$ &  $0.97 \pm 0.21$ \\ \hline \hline

 4 & 0.1  & $0.84 \pm 0.02$ & $0.93 \pm 0.05$ & $0.93 \pm 0.08$ & $0.97 \pm 0.08$ & $0.98 \pm 0.12$ & $0.96 \pm 0.07$ &  $1.00 \pm 0.17$ \\ \hline

 4 & -0.1  & $0.98 \pm 0.03$ & $0.96 \pm 0.08$ & $0.95 \pm 0.11$ & $0.99 \pm 0.12$ & $0.92 \pm 0.12$ & $0.95 \pm 0.13$ &  $0.92 \pm 0.15$ \\ \hline \hline

 5 & 0.1  & $0.81 \pm 0.04$ & $0.89 \pm 0.06$ & $0.90 \pm 0.05$ & $0.94 \pm 0.08$ & $0.96 \pm 0.11$ & $0.91 \pm 0.13$ &  $0.97 \pm 0.13$ \\ \hline

 5 & -0.1  & $0.98 \pm 0.02$ & $0.96 \pm 0.04$ & $0.99 \pm 0.10$ & $0.94 \pm 0.12$ & $0.92 \pm 0.07$ & $0.92 \pm 0.18$ &  $0.91 \pm 0.24$ \\ \hline \hline

 6 & 0.1  & $0.80 \pm 0.02$ & $0.88 \pm 0.03$ & $0.94 \pm 0.03$ & $0.95 \pm 0.08$ & $0.92 \pm 0.09$ & $0.91 \pm 0.11$ &  $0.99 \pm 0.13$ \\ \hline

 6 & -0.1  & $0.99 \pm 0.03$ & $0.97 \pm 0.07$ & $1.00 \pm 0.06$ & $1.01 \pm 0.07$ & $0.97 \pm 0.10$ & $0.97 \pm 0.11$ &  $0.93 \pm 0.14$ \\ \hline \hline

 7 & 0.1  & $0.75 \pm 0.02$ & $0.83 \pm 0.05$ & $0.90 \pm 0.06$ & $0.96 \pm 0.07$ & $0.94 \pm 0.08$ & $0.98 \pm 0.07$ &  $1.02 \pm 0.08$ \\ \hline

 7 & -0.1  & $0.99 \pm 0.04$ & $0.96 \pm 0.06$ & $0.95 \pm 0.07$ & $1.01 \pm 0.11$ & $1.01 \pm 0.13$ & $0.97 \pm 0.15$ &  $1.06 \pm 0.16$ \\ \hline \hline

\end{tabular}
\end{center}
\caption{\label{tab:on} Mean and $1 \sigma$ values of the scaling exponent $\alpha$ for variable $N$ and $\epsilon$ as a function of simulation time.}
\end{table} 

We have tested the veracity of these basic conclusions by performing a series of further simulations. As in the $N=1$ model we have tried a range of initial correlation sizes $c=1,2,4,8$. Again, it is found that the late time dynamics are independent of the initial conditions. We have also considered variations in the box size and have reduced the wall thickness by increasing $\epsilon$. For the most part the results are compatible with those presented in table~\ref{tab:on}.

\begin{figure}[] 
\centering
\mbox{\resizebox{0.8\textwidth}{!}{\includegraphics{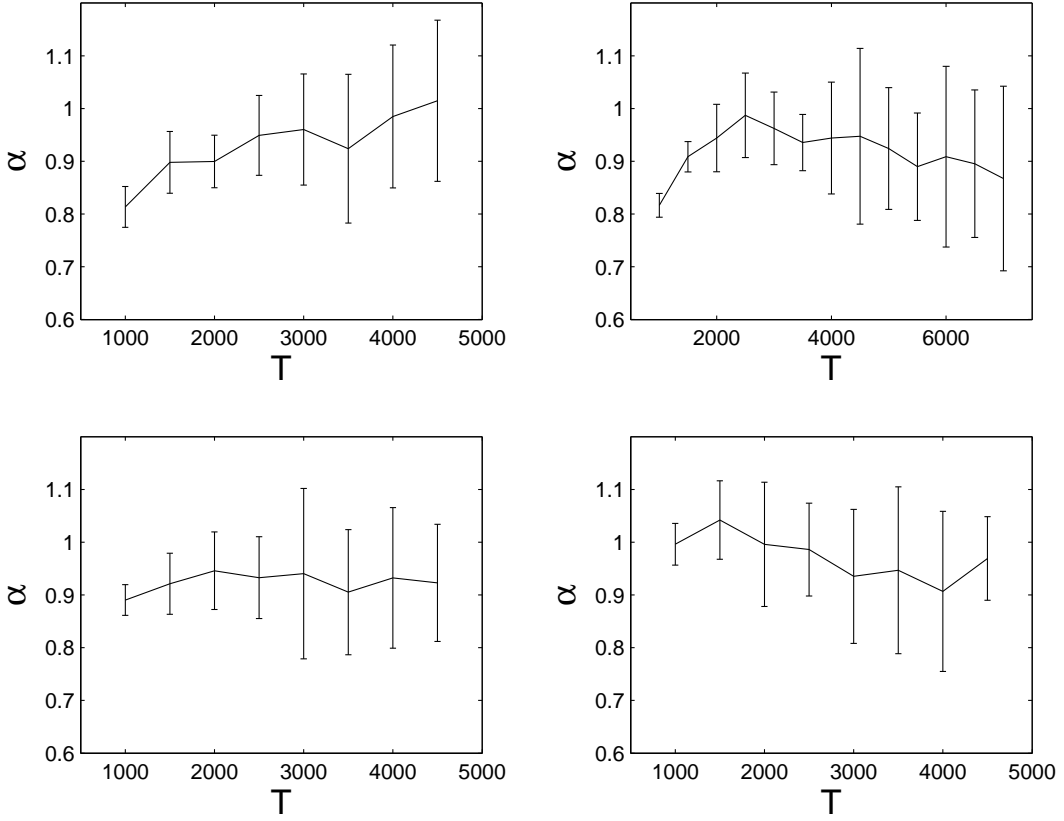}}}
\caption{Evolution of the scaling exponent $\alpha$ in the $N=5$ model. Top right as in table~\ref{tab:on}, top left using a $3000^{2}$ grid, bottom right with $\epsilon=0.5$ and bottom left using $\Delta x=0.8$, $\Delta t=0.2$. }
\label{alpha_o5big}
\end{figure}

We will concentrate our discussion on the case of $N=5$. We have performed runs with a larger grid of $3000^{2}$, an increased $\epsilon$ to 0.5 (which reduces the FWHM of the wall to 5 grid points as is the case for the $N=3$, $\epsilon=0.1$ model) and an increased value of $\Delta x=0.8$ (with $\Delta t$ adjusted to 0.2). The scaling exponents computed as a function of time are presented in Fig.~\ref{alpha_o5big}. Each one of these modifications increases the value of $\epsilon^{1/2}P\Delta x$ which is proportional to the size of the box, and hence the natural average wall separation, divided by the wall size. This is the figure of merit when discussing the effect of finite size effects on the exponents. These results would tend to suggest that one can explain the low values of $\alpha$ observed in the range $T=1000-2000$ for $N=5,6,7$ can be explained by a decrease in the ratio of the wall separation to the wall thickness.

\section{Discussion}

We have studied the stability and dynamical evolution of a network of domain walls where a global O(N) symmetry is broken to a (hyper)cubic symmetry group. In these models a variety of domain wall systems can arise with either X or Y type junctions. We have investigated the stability properties of hexagonal and square lattices arising from these junctions and have found results consistent with earlier work~\cite{BCM} using a continuum approximation. These results are encouraging in that they validate this method of assessing the stability of various wall configurations. However, they only show that these lattices are locally stable to small perturbations. They do not show that they are global attractor solutions.

In simulations where initial conditions are random, we find that the networks converge toward a self-similar scaling regime. There are hints that the scaling exponent is slightly less than 1, but we attribute these as being due to the finite box size and grid resolution which we are forced to use. These results are, in some sense, no surprise since the model has no scale in it which fixes the size of the domains. The only length scale is the horizon size and hence the wall network loses energy as fast as the causality allows, that is, $L\sim t$.

We conclude that the models considered here cannot give rise to a stable lattice from random initial conditions as would be required to explain the dark energy of the Universe. This does not, however, preclude the possibility of more complicated models containing junctions might lead to a lattice~\cite{nuno,carter}.

\section*{Acknowledgements} 

We have benefited from useful discussions with Nuno Antunes, Alan Bray, Brandon Carter, Elie Chachoua and Mark Hindmarsh.

\end{document}